\documentclass[numberedappendix,useAMS,usenatbib,letterpaper]{mn2e}
\usepackage[pass]{geometry}
\usepackage{amssymb,amsmath}
\usepackage{verbatim}
\usepackage{color,hyperref}
\usepackage{needspace}
\usepackage{enumitem}
\usepackage{tikz}
\usepackage{graphicx}
\usetikzlibrary{positioning}
\definecolor{linkcolor}{rgb}{0,0,0.25}
\hypersetup{
  colorlinks=true,        
  linkcolor=linkcolor,    
  citecolor=linkcolor,    
  filecolor=linkcolor,    
  urlcolor=linkcolor      
}
\newcounter{address}
\newcommand{\sectionname}{$\mathsection$}

\newcommand{\kpc}{\ensuremath{\,\mathrm{kpc}}}

\newcommand{\kms}{\ensuremath{\,\mathrm{km\ s}^{-1}}}

\newcommand{\inv}{\ensuremath{^{-1}}}

\title[Observing Hercules in APOGEE-2 South]
{The Hercules stream as seen by APOGEE-2 South}

\author[J. A. S. Hunt et al.]
{\parbox{\textwidth}{Jason A.~S.~Hunt$^1$, Jo~Bovy$^{1,2,3}$, Angeles P\'erez-Villegas$^{4}$, Jon A. Holtzman$^{5}$, \\Jennifer Sobeck$^{6}$, Drew Chojnowski$^{5}$, Felipe A. Santana$^{7}$, Pedro A. Palicio$^{8,9}$, Christopher Wegg$^{4}$, Ortwin Gerhard$^{4}$, Andr\'es Almeida$^{10}$, Dmitry Bizyaev$^{11,12}$, \\Jose G. Fernandez-Trincado$^{13,14}$, Richard R. Lane$^{15}$, Pen\'elope Longa-Pe\~na$^{16}$, \\Steven R. Majewski$^{17}$, Kaike Pan$^{11}$ and Alexandre Roman-Lopes$^{18}$}\vspace{0.5cm}
\\
$^{1}$ Dunlap Institute for Astronomy and Astrophysics, University of Toronto, 50 St. George Street, Toronto, Ontario, M5S 3H4, Canada\\
$^{2}$ Department of Astronomy and Astrophysics, University of Toronto, 50 St. George Street, Toronto, ON, M5S 3H4, Canada \\
$^{3}$ Alfred P. Sloan Fellow\\
$^{4}$ Max-Plank Institut f\"ur Extraterrestrische Physik, Gie{\ss}enbachstra{\ss}e, D-85741 Garching, Germany \\
$^{5}$ Department of Physics and Astronomy, New Mexico State University, Las Cruces, NM, 88003-8001, USA \\
$^{6}$ Department of Astronomy, Box 351580, University of Washington, Seattle, WA 98195, USA \\
$^{7}$ Universidad de Chile, Av. Libertador Bernardo O'Higgins 1058, Santiago de Chile \\
$^{8}$ Instituto de Astrof\'isica de Canarias, E-38205 La Laguna, Tenerife, Spain \\
$^{9}$ Universidad de La Laguna, Dpto. Astrof\'isica, E-38206 La Laguna, Tenerife, Spain \\
$^{10}$ Instituto de Investigaci\'on Multidisciplinario en Ciencia y Tecnolog\'ia, Universidad de La Serena, Benavente 980, La Serena, Chile \\
$^{11}$ Apache Point Observatory and New Mexico State University, P.O. Box 59, Sunspot, NM, 88349-0059, USA \\
$^{12}$ Sternberg Astronomical Institute, Moscow State University, Moscow \\
$^{13}$ Departamento de Astronom\'ia, Universidad de Concepci\'on, Casilla 160-C, Concepci\'on, Chile \\
$^{14}$ Institut Utinam, CNRS UMR6213, Univ. Bourgogne Franche-Comt\'e, OSU THETA, Observatoire de Besan\c{c}on, BP 1615, 25010 Besan\c{c}on Cedex \\
$^{15}$ Instituto de Astrofisica, Pontificia Universidad Cat\'olica de Chile, Av. Vicuna Mackenna 4860, 782-0436 Macul, Santiago, Chile \\
$^{16}$ Unidad de Astronom\'ia, Universidad de Antofagasta, Avenida Angamos 601, Antofagasta 1270300, Chile \\
$^{17}$ Department of Astronomy, University of Virginia, P.O. Box 400325, Charlottesville, VA 22904-4325, USA \\
$^{18}$ Departamento de F\'isica, Facultad de Ciencias, Universidad de La Serena, Cisternas 1200, La Serena, Chile \\
}

\pagerange{\pageref{firstpage}--\pageref{lastpage}}
\pubyear{2017}

\begin{document}

\maketitle

\label{firstpage}

\begin{abstract}
The Hercules stream is a group of co-moving stars in the Solar neighbourhood, which can potentially be explained as a signature of either the outer Lindblad resonance (OLR) of a fast Galactic bar or the corotation resonance of a slower bar. In either case, the feature should be present over a large area of the disc. With the recent commissioning of the APOGEE-2 Southern spectrograph we can search for the Hercules stream at $(l,b)=(270^\circ,0)$, a direction in which the Hercules stream, if caused by the bar's OLR, would be strong enough to be detected using only the line-of-sight velocities. We clearly detect a narrow, Hercules-like feature in the data that can be traced from the solar neighbourhood to a distance of about 4 kpc. The detected feature matches well the line-of-sight velocity distribution from the fast-bar (OLR) model. Confronting the data with a model where the Hercules stream is caused by the corotation resonance of a slower bar leads to a poorer match, as the corotation model does not predict clearly separated modes, possibly because the slow-bar model is too hot.\end{abstract}


\section{Introduction}\label{sec:intro}
It has been known for considerable time that the Milky Way is a barred galaxy, initially inferred from observations of gas kinematics \citep[e.g.][]{Vaucouleurs64}, and later confirmed by infrared photometry \citep[e.g.][]{BS91}. However, some of the fundamental parameters of the bar have not been fully determined, in particular its length, mass, pattern speed and angle with respect to the Sun--Galactic-center line remain the subject of debate. This is primarily owing to the high levels of dust extinction present in the Galactic plane, especially when looking towards the inner Galaxy, and because the structure and kinematics of the bar and/or bulge region is highly complex \citep[e.g.][]{C-LG-FGHL-C08,Nataf+10,McWZ10,Ness+16,Williams+16,Portail+17}.

However, as an alternative to direct observation, we are able to infer properties of the bar from its effect on the orbits of stars further out in the Galaxy, especially those strongly affected by resonant interactions. Specifically, a rotating galactic bar has three fundamental resonances; the corotation resonance (CR), where $\Omega_\phi = \Omega_{\text{b}}$, and the inner (ILR) and outer (OLR) Lindblad resonances such that
\begin{equation}
\Omega_{\text{b}}=\Omega_{\phi}\pm\frac{1}{2}\Omega_{\text{R}},
\end{equation}
where $\Omega_{\text{b}}$ is the pattern speed of the bar and $\Omega_{\phi}$ and $\Omega_{\text{R}}$ are the tangential and radial orbital frequencies. \cite{D00} proposed that the Hercules stream, a well known feature in the kinematics of the Solar neighbourhood, is caused by the OLR of the bar. To reproduce the observed kinematics, this explanation requires the OLR to be just inside the Solar circle, with a bar pattern speed of around 1.85 times the local circular frequency. Note that while there are other possible explanations for Hercules, e.g. a moving group with a common birthplace within the disc, or the signature of a merger, the fact that it is comprised of old and late type stars \citep[e.g.][]{Dehnen98} and spans a range of ages and metallicities \citep[e.g.][]{2005A&A...430..165F,2007ApJ...655L..89B,BH10} argues for a dynamical origin. For this explanation \cite{D00} found the constraint $\Omega_{\text{b}}=(1.85\pm0.05)\times\Omega_0$, or  $\Omega_{\text{b}}=53\pm3$ km s$^{-1}$ kpc$^{-1}$ for $\Omega_0=28.5$ km s$^{-1}$ kpc$^{-1}$ This is in agreement with later work. For example, \cite{MNQ07} find $\Omega_{\text{b}}=(1.87\pm0.04)\times\Omega_0$ with an analysis of the Oort constants, and \cite{MBSB10} find $\Omega_{\text{b}}=(1.82\pm0.07)\times\Omega_0$.

\cite{Aetal14-2} use data from the Radial Velocity Experiment \citep[RAVE;][]{Sea06} to trace Hercules between Galactic radii $R=7.8$ to 8.6 kpc (assuming the Sun to be located at $R_0=8.05$ kpc). They find best fitting values of $\Omega_{\text{b}}=(1.89\pm0.08)\times\Omega_0$, or  $\Omega_{\text{b}}=56\pm2$ km s$^{-1}$ kpc$^{-1}$ for $\Omega_0=29.5$ km s$^{-1}$ kpc$^{-1}$ under the assumption that the Hercules stream is caused by the OLR. \cite{Monari+16} use data from the Large Sky Area Multi-Object Fibre Spectroscopic Telescope \citep[LAMOST;][]{ZZCJD12} combined with data from the Tycho-$Gaia$ Astrometric Solution \citep[TGAS;][]{GaiaDR1} to trace the Hercules stream in the direction of the Galactic anticentre, up to a distance of 0.8 kpc. They find good agreement with the values from \cite{Aetal14-2}. These works show that the RAVE and LAMOST data are consistent with Hercules being caused by the OLR of the bar, and the measured pattern speeds put an upper limit on the length of the bar, which cannot extend past corotation \citep[e.g.][]{Contopoulos80}. If Hercules is caused by the OLR, the bar is expected to end around 3--3.5 kpc from the Galactic centre.

\begin{figure}
\centering
\includegraphics[width=\hsize]{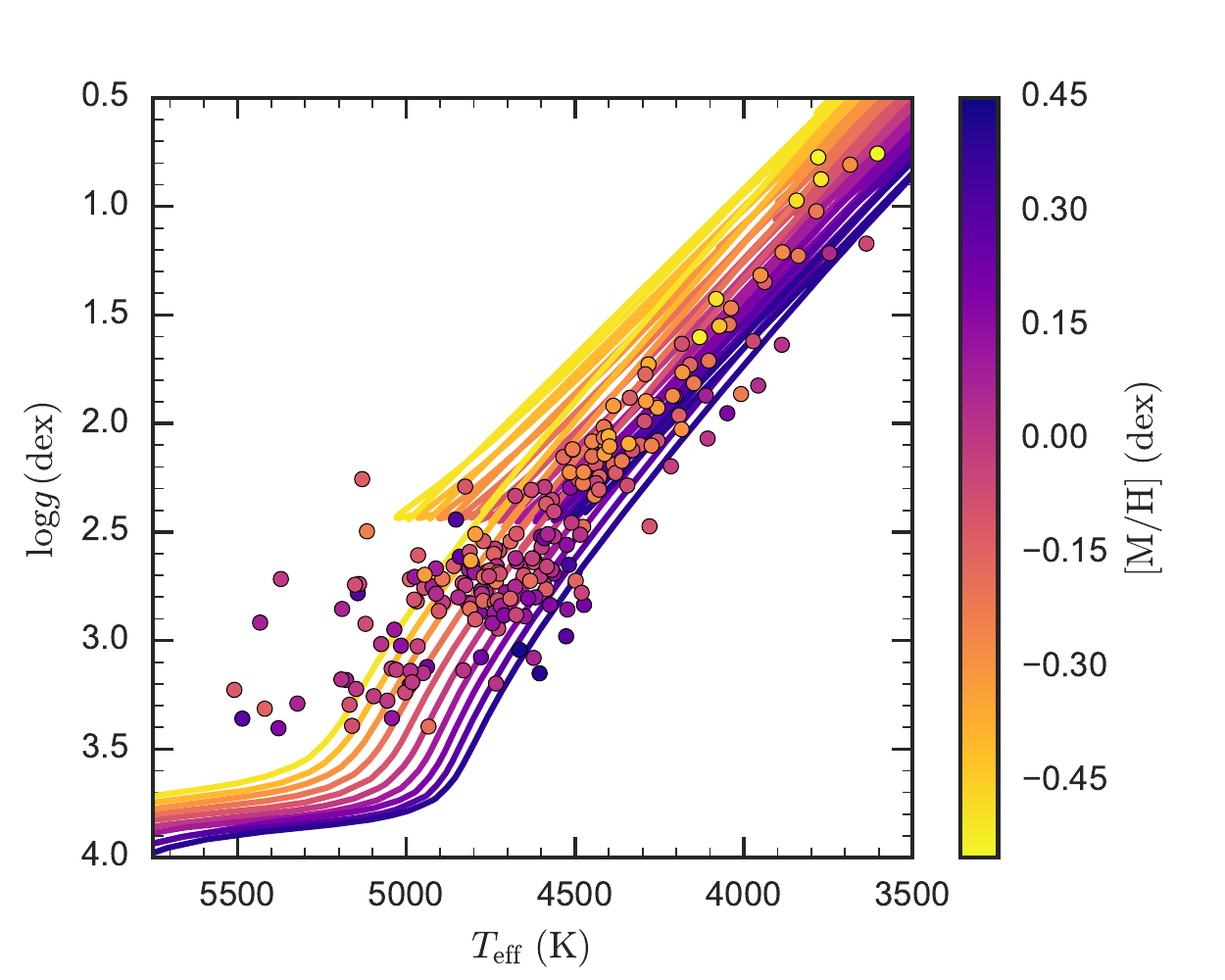}
\caption{$T_{\mathrm{eff}}$ vs. $\log g$ colour-coded by overall metallicity for the giant stars in the $(l,b) = (270^\circ,0)$ field, overlaid on a small number of PARSEC isochrones for an age of 5 Gyr, colour-coded by the same metallicity scale.}
\label{tefflogg}
\end{figure}

However, some recent studies of the bar favour a notably longer \citep[e.g.][]{WGP15} and slower bar \citep[e.g.][]{2017MNRAS.465.1621P}. This is at odds with the interpretation of Hercules as a resonance feature of the OLR because in this scenario the OLR is {a few kpc} outside the Solar circle. For example, in \cite{P-VPWG17}, for a bar with half length $\sim5$ kpc and pattern speed $\Omega_{\text{b}}=39$ km s$^{-1}$ kpc$^{-1}$, the OLR is located at $R\sim10.5$ kpc, and cannot be responsible for the Hercules stream. However, it is worth noting that an OLR like feature has been observed by \cite{Liu+12} at $10-11$ kpc, which is consistent with this model. If the bar does extend to $\sim5$ kpc, then an alternate explanation is needed for the Hercules stream. 

\cite{P-VPWG17} show that a 5 kpc bar in an $N$-body model is also able to reproduce the Hercules stream in the Solar neighbourhood via a different mechanism, where stars orbiting the bar's Lagrange points $L_4$ and $L_5$ move outwards from corotation, which occurs at $R=6$ kpc, and reach the Solar neighbourhood. In this model, the Hercules stream will be stronger within the Solar radius, and should disappear a few hundred parsecs further from the centre. This is in contrast with the fast-bar model, where the Hercules stream should extend further beyond the Solar radius \citep[e.g.][]{Bovy10}. In the near future, the European Space Agency's $Gaia$ mission \citep{GaiaMission} will enable us to probe much further from the Solar neighbourhood, and potentially trace the extent of Hercules, determining which resonance is responsible. However, \cite{Bovy10} showed that in specific directions ($250^{\circ} < l < 290^{\circ}$), line-of-sight velocities alone are enough to observe Hercules if it is caused by the OLR, because in these directions the Hercules feature is well separated from the main mode of the velocity distribution along the direction that projects onto the line of sight.

The Sloan Digital Sky Survey IV \citep[SDSS-IV; e.g.][]{SDSS4}, Apache Point Observatory Galactic Evolution Experiment \citep[APOGEE;][]{MAPOGEE16,MAPOGEE17} has provided high resolution spectroscopic data for around 150,000 stars in the Northern hemisphere during its first phase, APOGEE-1 \citep{Holtzman15a}. Now, as part of its second phase, APOGEE-2, which includes a second spectrograph in the Southern hemisphere, APOGEE-2S, will extend APOGEE's coverage across the sky. The line-of-sight $(l,b)=(270^\circ,0)$ is one of the first fields observed during the commissioning of the APOGEE-2S spectrograph, which enables us to look for the Hercules stream at the line-of-sight identified in \cite{Bovy10} as having a strong feature.

This paper is constructed as follows. In \sectionname~\ref{observation} we discuss our treatment of the data and the resulting velocity distribution. In \sectionname~ \ref{modeling} we present the model predictions for this line of sight from both the fast-bar model, and the slow-bar model from \cite{P-VPWG17}. In \sectionname~ \ref{comparison} we compare the models with the data. In \sectionname~ \ref{conclusion} we discuss the implications of the detection and look forward to future data.

\section{Observed feature}\label{observation}

\begin{figure}
\centering
\includegraphics[width=\hsize]{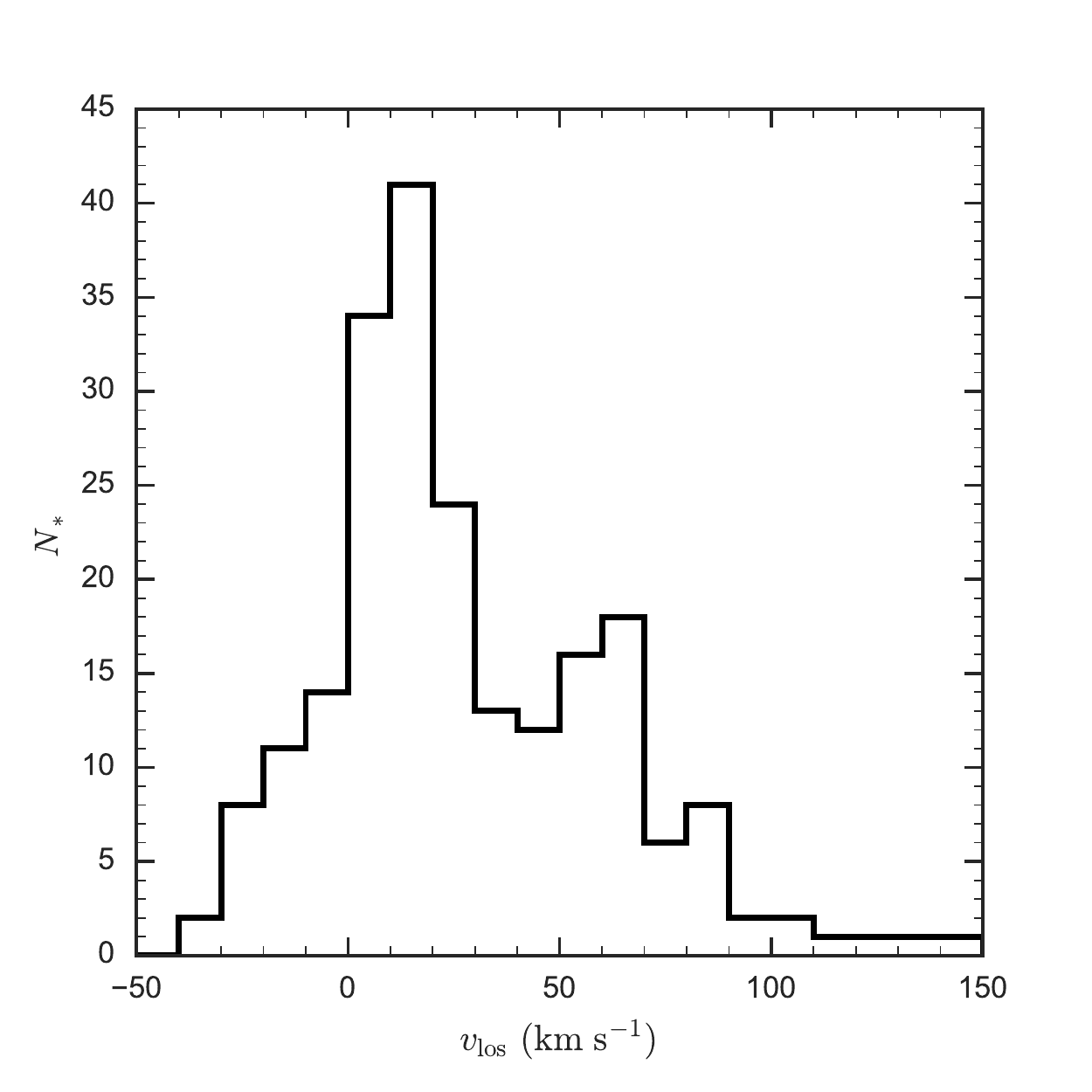}
\caption{Histogram of line-of-sight for stars in the $(l,b)=(270^\circ,0)$ APOGEE-2S field. The distribution displays a clear secondary peak.}
\label{hist}
\end{figure}

We use spectroscopic data from the field centered on $(l,b) =
(270^\circ,0)$ taken during commissioning of the APOGEE-2 Southern
spectrograph installed at the 2.5 m du Pont telescope \citep{duPont}
at Las Campanas Observatory. This spectrograph is almost an exact
clone of the APOGEE spectrograph installed at the 2.5 m Sloan
Foundation telescope \citep{2006AJ....131.2332G} and described in
(\citealt{2010SPIE.7735E..1CW}; J. Wilson et al. 2017, in
preparation). The target selection is described in \cite{Zasowski+17}. A fully customised reduction pipeline for the APOGEE-2S
spectrograph is not yet available and we have therefore processed the
spectra using the existing APOGEE reduction
\citep{2015AJ....150..173N} and stellar-parameter
\citep[ASPCAP;][]{2016AJ....151..144G} pipelines. Like the original APOGEE
spectrograph, the APOGEE-2S spectrograph has 300 fibers and the data
for the $(l,b) = (270^\circ,0)$ field contain 262 stars (with 35 fibers assigned to observe the sky, 2 unmapped fibers, and one with bad pixels). 

Because this is the first application of the existing pipeline to data
from the new spectrograph, we examine all the spectra and their radial velocity
template fits by eye to remove low signal-to-noise ratio spectra and prevent
bad template matching. We then run these spectra through the reduction
and stellar-parameters pipeline to determine line-of-sight velocities
$v_{\text{los}}$, effective temperature $T_\mathrm{eff}$, surface
gravity $\log g$, and the overall metallicity $[\mathrm{M/H}]$. Other parameters ($\mathrm{[C/M]}$, $\mathrm{[N/M]}$, $\mathrm{[\alpha/M]}$, $\xi$) are simultaneously fit by the pipeline, but we will ignore them for the time being. Our main use of the stellar parameters is to infer distances. 

We correct the line-of-sight velocities using the solar motion from
\citep{Bovy+12,2015ApJ...800...83B}; because the observed field is
small, we apply this correction as a constant correction of $-24\kms$
and $v_{\text{los}}$ from here on refers to the corrected
value. Note that toward $(l,b) = (270^\circ,0)$ only the tangential
  component $V_\odot$ of the Solar motion matters. Applying a
  different Solar motion would simply shift all of the line-of-sight
  velocities by a constant without changing the shape of the
  distribution of line-of-sight velocities. 
  
  We then select giants by
requiring $\log g \leq3.5$ and only consider stars with $-200\leq
v_{\text{los}}\leq 200$ km s$^{-1}$ to remove outliers and possible
halo stars. The remaining sample consists of 218 giant
stars. Fig. \ref{tefflogg} shows $T_{\text{eff}}$ versus $\log g$
colour-coded by metallicity [M/H] for these 218 stars, overlaid on a small number of PARSEC isochrones \citep{Bressan+12} for an age of 5 Gyr, for illustration. Even though the data were only run through preliminary versions of the reduction
pipelines, the stellar parameters follow the expected behavior as a
function of $(T_\mathrm{eff},\log g,[\mathrm{M/H}])$ along the giant
branch. Note that the full calibrations are not yet available for APOGEE-2S, thus this version of the pipeline uses the raw ASPCAP output, which produces reliable [M/H] and $\mathrm{T_{eff}}$, but can show an offset in $\mathrm{log}g$ of $0.2-0.3$ dex relative to the asteroseismic gravities for giants.

\begin{figure}
\centering
\includegraphics[width=\hsize]{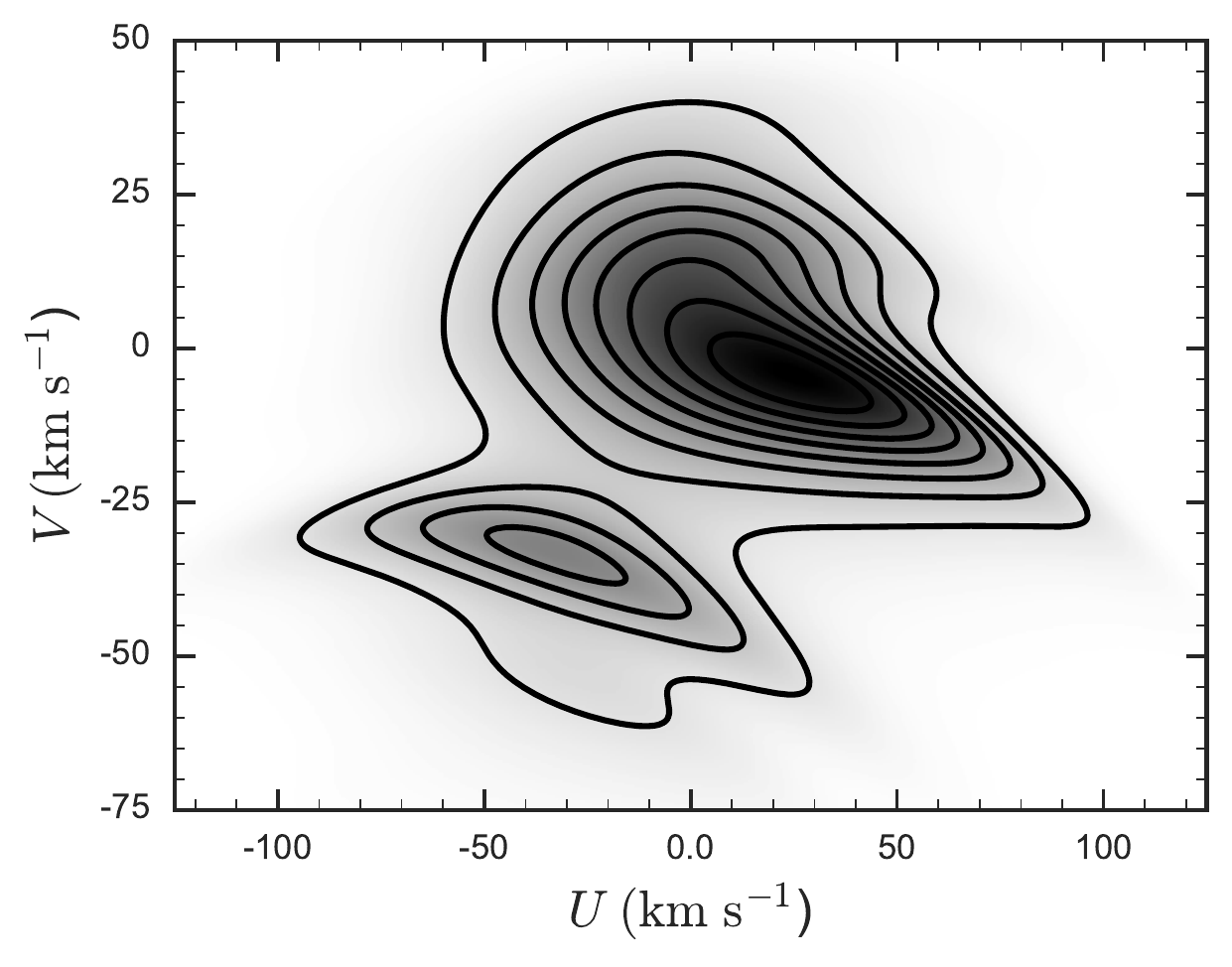}
\caption{The velocity distribution in the Solar neighbourhood for the fast-bar model from \sectionname~\ref{sec:fastbar} for the chosen model parameters.}
\label{uvmodel}
\end{figure}

Fig. \ref{hist} shows the distribution of line-of-sight velocities for the full sample. This distribution displays a clearly bimodal distribution. This structure is qualitatively what we expect to see for the Hercules stream, which is located around $V=-50$ km s$^{-1}$ in the Solar neighbourhood. The data prefer a double Gaussian fit over a single Gaussian fit at a significance of $4.8\sigma$, although note that neither mode is expected to be Gaussian in nature.

We also wish to test how the Hercules feature changes with distance from the Sun. To split our sample in different distance bins, we estimate distances for the stars in our sample using the isochrone fitting code \sc{unidam }\rm \citep{MH17}. \sc{unidam }\rm estimates distances from the available stellar parameters ($T_{\text{eff}}$, $\log g$, and [M/H]), and from the $J$, $H$, $K_s$ magnitudes, by comparing the data with PARSEC 1.2S isochrones in a Bayesian manner. \sc{unidam }\rm fails for 3 stars not already excluded by earlier cuts, providing a final sample of 215 stars with distance estimates, with a median distance uncertainty of 357 pc. This may be underestimated owing to the uncertainty in $\mathrm{log}g$. However, \cite{Bovy10} shows that the secondary peak of Hercules remains clearly visible with distance errors of $30\%$, so the relative distance error of $10-20\%$ expected here should be sufficient for this analysis.

\section{Expectation from Galactic models}\label{modeling}

To compare the data to the competing resonance models discussed in the
introduction, we make predictions for the observed line-of-sight
velocity distribution in this section. In
\sectionname~\ref{sec:fastbar}, we describe how we make detailed
predictions using the formalism of \citet{D00} and \citet{Bovy10} for
the model where the Hercules feature represents the effect of the
OLR. In \sectionname~\ref{sec:slowbar}, we discuss how we extract
predictions for the slow-bar model from the work of \citet{P-VPWG17}.

\subsection{The fast bar}\label{sec:fastbar}

To make predictions for the fast-bar model, which is capable of reproducing the Hercules stream in the Solar neighbourhood through the OLR \citep{D00}, we use \texttt{galpy}\footnote{Available at \url{https://github.com/jobovy/galpy}~.} \citep{B15} to simulate the orbits of stars along the $(l,b)=(270^\circ,0)$ line-of-sight. Because all of the stars in our sample are close to the Galactic mid-plane, we only simulate the two-dimensional dynamics in the Galactic plane, ignoring vertical motions. To represent the distribution of stellar orbits, we use a Dehnen distribution function \citep{Dehnen99} to model the stellar disc before bar formation. This distribution function is a function of energy $E$ and angular momentum $L$
\begin{equation}
f_{\text{dehnen}}(E,L)\ \propto \frac{\Sigma(R_e)}{\sigma^2_{\text{R}}(R_e)}\exp\biggl[\frac{\Omega(R_e)[L-L_c(E)]}{\sigma^2_{\text{R}}(R_e)}\biggr],
\end{equation}
where $R_e$, $\Omega(R_e)$, and $L_c$ are the radius, angular frequency, and angular momentum, respectively, of the circular orbit with energy $E$. The gravitational potential is assumed to be a simple power-law, such that the circular velocity is given by
\begin{equation}
  v_c(R)=v_0(R/R_0)^{\beta}\,,
\end{equation}
where $v_0$ is the circular velocity at the solar circle at radius $R_0$. To model the bar we use the simple quadrupole bar potential from \citet{D00} given by
\begin{equation}
\begin{split}
\Phi(R,\phi)&=A_{\text{b}}(t)\cos(2(\phi-\Omega_{\text{b}}t))\\ 
& \quad \quad \times
\left\{ \begin{array}{ll} -(R_{\text{b}}/R)^3, & \mathrm{for}\ R \geq R_{\text{b}},\\ (R/R_{\text{b}})^3-2, & \mathrm{for}\ R \leq R_{\text{b}}. \end{array}
\right.
\end{split}
\end{equation}
where $R_{\text{b}}$ is the bar radius, set to $80\%$ of the corotation radius. The bar is grown smoothly following the prescription
\begin{eqnarray}
A_{\text{b}}(t)=
\left\{\begin{array}{ll} 0,\ \frac{t}{T_{\text{b}}}<t_{\text{1}} \\ A_f\biggl[\frac{3}{16}\xi^5-\frac{5}{8}\xi^3+\frac{15}{16}\xi+\frac{1}{2}\biggr], t_{\text{1}}\leq\frac{t}{T_{\text{b}}}\leq t_{\text{1}}+t_{\text{2}}, \\ A_f,\ \frac{t}{T_{\text{b}}} > t_{\text{1}} + t_{\text{2}}.  \end{array}
\right.\,
\end{eqnarray}
where $t_1$ is the start of bar growth, set to half the integration time, and $t_2$ is the duration of the bar growth. $T_{\text{b}}=2\pi/\Omega_{\text{b}}$ is the period of the bar,
\begin{equation}
\xi=2\frac{t/T_{\text{b}}-t_{\text{1}}}{t_{\text{2}}}-1,
\end{equation}
and
\begin{equation}
A_f=\alpha\frac{v_0^2}{3}\biggl(\frac{R_0}{R_{\text{b}}}\biggr)^3,
\end{equation}
where $\alpha$ is the dimensionless strength of the bar. For our fast bar model, we set $\alpha=0.01$, $R_0=8.0$ kpc, and $v_0=220$ km s$^{-1}$.

Following \cite{D00} we specify the final time $t_2$, and integrate backwards to $t=0$. This is valid because the potential is known, and the distribution function (DF) remains constant along stellar orbits, as known from the collisionless Boltzman equation. Thus the value of the DF $f(\bold{r},t_2)$ for some phase space coordinates $\bold{r}=(\bold{x}, \bold{v})$ at $t_2$ is equal to $f(\bold{r}_0,0)$ if $\bold{r}$ is the result of integrating $\bold{r}_0$ from $t=0$ to $t=t_2$. Therefore we can obtain $\bold{r}_0$ by integrating the orbits backwards in time.

\begin{figure}
\centering
\includegraphics[width=\hsize]{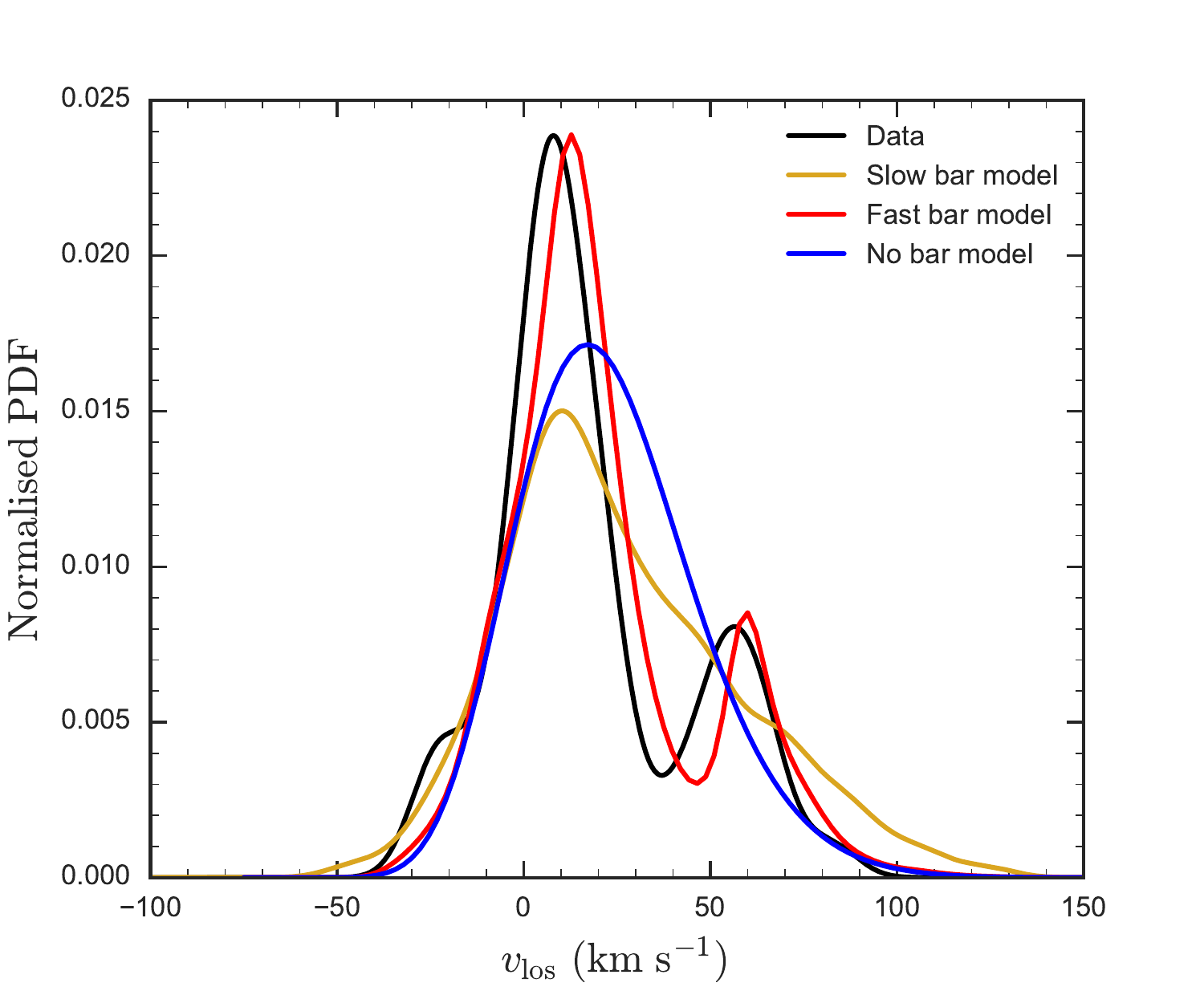}
\caption{KDE of the data in the range $2.2\leq d\leq3.2$ kpc (black) overlaid with the axisymmetric prediction (blue), the fast-bar model prediction (red), and the slow-bar model (gold).}
\label{kde}
\end{figure}

The fiducial model from \citet{D00} and that explored in detail by
\citet{Bovy10} had a flat rotation curve ($\beta = 0$) and OLR radius
of $R_{\text{OLR}}=0.9\,R_0$. We slightly adjust these parameters to
provide a better match to the data in \sectionname~\ref{comparison}:
$R_{\text{OLR}}=0.95\,R_0$ and $\beta=-0.1$ (a slightly falling
rotation curve). This change causes the separation of the Hercules
feature from the main mode of the velocity distribution to be slightly
smaller. We combine this gravitational-potential model with a Dehnen
distribution function appropriate for the kinematics of the APOGEE
sample of intermediate-age stars \citep{Bovy+12}: surface-density
scale length $R_0/3$ and an exponential radial velocity dispersion
with scale length $R_0$ and $\sigma_R(R_0)=33\kms$. The bar has an angle of $25^{\circ}$  with respect to the Sun--Galactic-center line, a pattern speed of $\Omega_{\text{b}}=48.6$ km s$^{-1}$ kpc$^{-1}$ and co-rotation occurs at $R_{\text{CR}}=4.77$ kpc.

Fig. \ref{uvmodel}
displays the local velocity $(U,V)$ distribution in the Solar
neighbourhood for our chosen bar parameters, where $U$ is the velocity in the direction of the Galactic centre, and $V$ is the velocity in the direction of rotation with respect to the local standard of rest (LSR). It clearly reproduces a Hercules-like bimodality, with the density peak of the Hercules feature occurring around $U=-30$ km s$^{-1}$ with respect to the LSR, or $U=-40$ km s$^{-1}$ with respect to the Sun, which is in agreement with the literature \citep[e.g.][]{Dehnen98,2005A&A...430..165F,BHR09,P-VPWG17}. We do not expect this simple model to well reproduce the main density peak of the Solar neighborhood, which is heavily influenced by other non-axisymmetric components and dynamical processes such as the Spiral structure and other moving groups.

Below we also compare the data to a simple axisymmetric model. This
model is that of this section, except that it has no bar.

\subsection{The slow bar}\label{sec:slowbar}

To make predictions for the slow bar, which can reproduce a Hercules-like feature in the velocity distribution in the Solar neighbourhood through orbits around the bar's Lagrange points, we examine the velocity distribution of the model from \cite{P-VPWG17}. We do this rather than evaluating a simple model like that for the fast bar described above, because simple models for the slow bar do not reproduce the Hercules stream in the solar neighbourhood \citep{2017MNRAS.465.1443M}. The model is based on the Made-to-Measure \citep[M2M; e.g.][]{ST96,DeL07,HK13} model from \cite{2017MNRAS.465.1621P} which reproduces a range of observational constraints in the bulge and bar region. In this model, the bar has a pattern speed of $\Omega_{\mathrm{b}}=39\kms\kpc\inv$, with an angle of $28^{\circ}$  with respect to the Sun--Galactic-center line. The Sun is located at $R_0=8.2$ kpc, and the local circular velocity $v_c(R_0)=243$ km s$^{-1}$.

The effective potential of the bar is constrained by the fitted bulge-bar data, and therefore the perturbations and resonant orbits in the outer disc are predicted. However the population of these orbits and the distribution function in the outer disc is not fitted. \cite{P-VPWG17} adjust the model from \cite{2017MNRAS.465.1621P} by increasing the resolution by a factor of 10, following a variation of the resampling scheme presented in \cite{Deh09}. Because no data was fitted in the outer disc of this model, the velocity dispersion of the disc was unconstrained and depends heavily on the initial N-body model which was hotter than the Milky Way's disc. Therefore \cite{P-VPWG17} cooled the outer disc to $\sigma_{\mathrm{R}}=35.5$ km s$^{-1}$ to reproduce the local radial velocity dispersion. However it is clear from  Figure 2 of that work that even this cooled disc is too hot to reproduce the rich and complex substructure in the $UV$ diagram. Fitting the model to this local velocity distribution is currently infeasible because it would require knowledge of the presently ill constrained potential of the spiral arms. Therefore we compare the model of \cite{P-VPWG17} as it is, selecting particles in a 100 pc cylinder along the $(l,b)=(270^\circ,0)$ line-of-sight assuming $R_0=8.2$ kpc following the model.

\begin{figure*}
\centering
\includegraphics[width=\hsize]{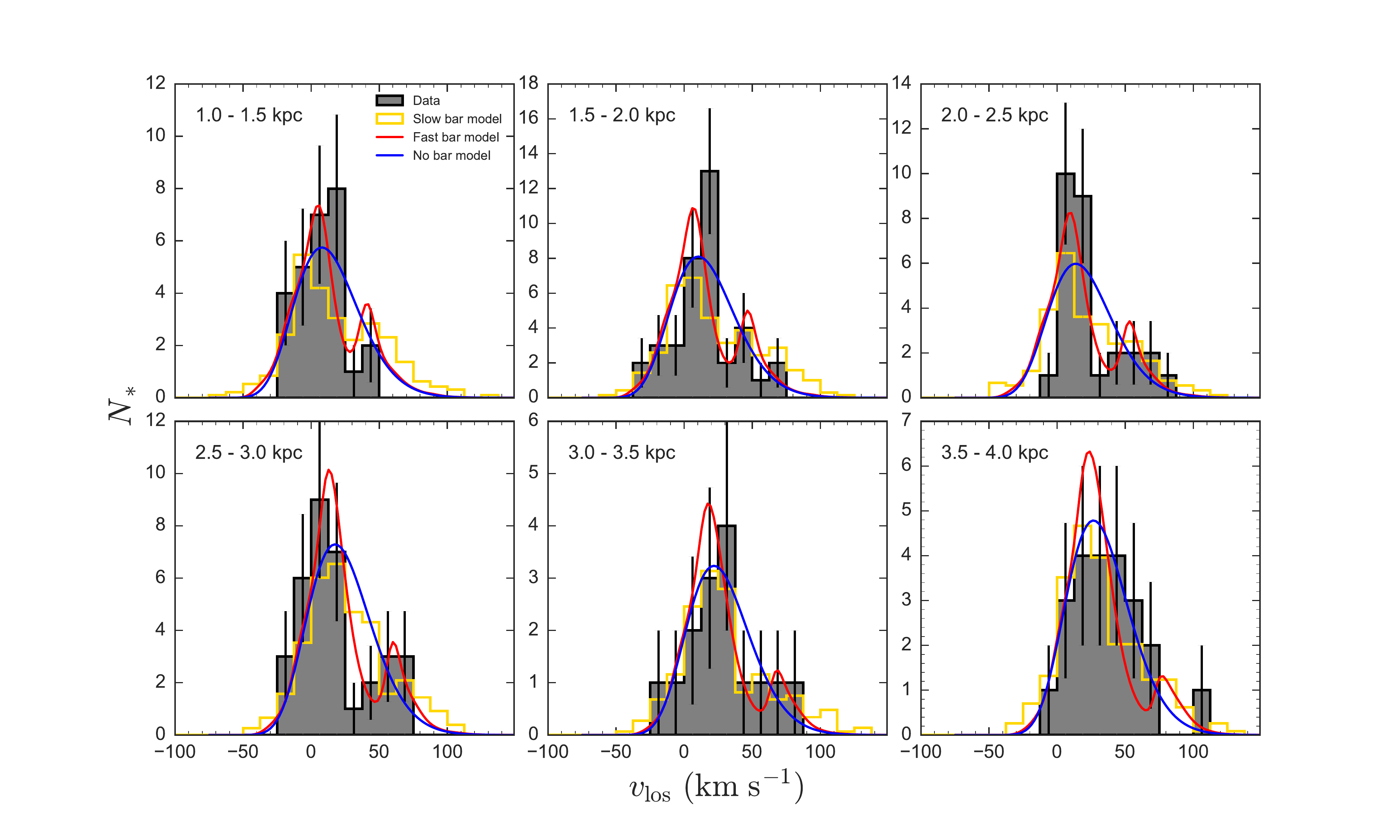}
\caption{Histogram of the line-of-sight velocity distribution (km s$^{-1}$) for the data (black) overlaid with the prediction from an axisymmetric model (blue), the fast-bar model (red), and the slow-bar model (gold) for six distance bins in the range $1-4$ kpc. The error bars show the Poisson errors for the data. All three models have been normalised to match the number counts in the data.}
\label{vlos}
\end{figure*}

\section{Comparison with data}
\label{comparison}

Fig. \ref{kde} shows a Kernel Density Estimate \citep[KDE;][]{W06} with bandwidth 7 km s$^{-1}$ of the observed line-of-sight velocity distribution (black) in the range $2.2\leq d\leq3.2$ kpc, chosen for being the range where the feature appears to be strongest in the data. We compare the data distribution to the prediction from the short bar model (red) and a KDE prediction from slow-bar model both computed at the same distance. We also include the prediction from a simple axisymmetric for reference; this is not expected to reproduce the Hercules peak. 

Note that \cite{P-VPWG17} assume a Solar motion of 12.24 $\kms$ \citep{SBD10}. We remind the reader that in this work we compare the models and data in the local standard of rest frame having corrected the data by $24$ $\kms$. A different Solar motion would shift the observed velocity distribution without changing its shape.

There is excellent agreement between the fast-bar model prediction and the data, except for a 4 km s$^{-1}$ misalignment of the peaks, which could be due to a slight overestimation of the Solar velocity. The slow-bar model is clearly too hot compared to the data. This is already known from the comparison to the Solar neighborhood $UV$ plot in \cite{P-VPWG17}; the larger width of the main peak is owing to the high dispersion, not the orbital origin of Hercules. The slow-bar model does not have a clear Hercules-like peak, unlike the data and the fast-bar model. However, note that the model was not fitted to reproduce the velocity distribution in the outer disc, aside from cooling it to better reproduce the local velocity dispersion. It was fitted to the entire bar region, which naturally produces a feature consistent with Hercules in the Solar neighborhood.

Fig. \ref{vlos} displays the histogram of the line-of-sight velocity distribution for the data in black for six 500 pc wide distance bins between 1 and 4 kpc. These histograms show that a clear second peak is present in the data in the three bins between $1.5$ and $3\kpc$. We also overlay the predictions for the axisymmetric, fast-bar, and slow-bar models. For the distance bins under 3 kpc, the fast-bar model shows an excellent match to the data. The distance bins over 3 kpc have comparatively few stars, and it is hard to distinguish between the axisymmetric and both bar models which all fit the data well. Qualitatively, the slow-bar model can populate the area of the distribution containing the Hercules stream. However, there is no clear division between the central peak and Hercules in the slow-bar model. It is difficult to say whether this is an intrinsic feature of the mechanism which populates the orbit space, or a result of the high velocity dispersion.

\section{Discussion and outlook}\label{conclusion}

We have shown that there is a bimodality in the line-of-sight velocity distribution in data taken by APOGEE-2S along $(l,b)=(270^\circ,0)$ up to around 4 kpc, matching the expected location of the Hercules stream. This is substantially further from the Sun than any existing detection of the Hercules stream. It leaves no doubt that the Hercules stream is indeed a large-scale dynamical feature in the Milky Way, rather than a localized solar-neighbourhood anomaly.

The stream can plausibly be explained as the effect of OLR of a short, fast-rotating Galactic bar \citep{D00}, or as the result of stars orbiting the Lagrange points of a long slower-rotating Galactic bar \citep{P-VPWG17}. We compared the observed velocity distribution with predictions from the two models around a distance of $2.7\kpc$ where the feature was strongest in the data. We find that the fast-bar model is an excellent match to the data, while the slow-bar model does not reproduce two clearly divided modes. We also compared the data with the models in six separate distance bins from 1 to 4 kpc. Again we find that the fast-bar model is a better match. The slow-bar model consistently populates the area of the distribution containing the Hercules stream, but without the observed bimodal structure. This could be owing to the high velocity dispersion obscuring any separation between the modes, and thus future work is necessary to better predict the expected velocity distribution in the slow-bar model for the APOGEE sample.

We also note here that we do not find that the stars in the Hercules stream have systematically higher metallicity than the sample in general. This is in agreement with $\mbox{\cite{BH10}}$, who find no significant metallicity preference in Hercules in the Solar neighbourhood, but in conflict with \cite{RGMFU98} who report the anomaly in the $(U,V)$ plane (Hercules) to consist of intermediate to high metallicity stars.

In the near future, $Gaia$ will provide 5D phase space information for over $10^9$ stars down to $\sim20$ mag, and radial velocities for around $10^8$ stars down to $\sim16$ mag. The brightest stellar types will be visible across many kpc, enabling us to trace the extent of Hercules in all directions and also refine our measurements of the extent of the Galactic bar. At this stage it should become clear whether the Hercules stream conforms to the OLR explanation for its origin, or whether another mechanism is required to explain the observed velocity distribution across the disc.

\section*{Acknowledgements} We thank the anonymous referee for their valuable comments. We would like to thank Alexey Mints for valuable help with the \sc{unidam }\rm code. JASH is supported by a Dunlap Fellowship at the Dunlap Institute for Astronomy \& Astrophysics, funded through an
endowment established by the Dunlap family and the University of
Toronto. JB received partial support from the Natural Sciences and
Engineering Research Council of Canada. JB also received partial
support from an Alfred P. Sloan Fellowship. RRL acknowledges support by the Chilean Ministry of Economy, Development,
and Tourism's Millennium Science Initiative through grant IC120009,
awarded to The Millennium Institute of Astrophysics (MAS). RRL also
acknowledges support from the STFC/Newton Fund ST/M007995/1 and the
CONICYT/Newton Fund DPI20140114. PLP was supported by MINEDUC-UA project, code ANT 1655.

Funding for the Sloan Digital Sky Survey IV has been provided by
the Alfred P. Sloan Foundation, the U.S. Department of Energy Office of
Science, and the Participating Institutions. SDSS-IV acknowledges
support and resources from the Center for High-Performance Computing at
the University of Utah. The SDSS web site is www.sdss.org.

SDSS-IV is managed by the Astrophysical Research Consortium for the 
Participating Institutions of the SDSS Collaboration including the 
Brazilian Participation Group, the Carnegie Institution for Science, 
Carnegie Mellon University, the Chilean Participation Group, the French Participation Group, Harvard-Smithsonian Center for Astrophysics, 
Instituto de Astrof\'isica de Canarias, The Johns Hopkins University, 
Kavli Institute for the Physics and Mathematics of the Universe (IPMU) / 
University of Tokyo, Lawrence Berkeley National Laboratory, 
Leibniz Institut f\"ur Astrophysik Potsdam (AIP),  
Max-Planck-Institut f\"ur Astronomie (MPIA Heidelberg), 
Max-Planck-Institut f\"ur Astrophysik (MPA Garching), 
Max-Planck-Institut f\"ur Extraterrestrische Physik (MPE), 
National Astronomical Observatories of China, New Mexico State University, 
New York University, University of Notre Dame, 
Observat\'ario Nacional / MCTI, The Ohio State University, 
Pennsylvania State University, Shanghai Astronomical Observatory, 
United Kingdom Participation Group,
Universidad Nacional Aut\'onoma de M\'exico, University of Arizona, 
University of Colorado Boulder, University of Oxford, University of Portsmouth, 
University of Utah, University of Virginia, University of Washington, University of Wisconsin, 
Vanderbilt University, and Yale University.

\bibliographystyle{mn2e}
\bibliography{ref2}

\end{document}